\documentclass[twocolumn,aps, prl]{revtex4}

\usepackage{graphics}
\usepackage{graphicx}
\usepackage{dcolumn}
\usepackage{bm}
\usepackage{amsmath,amssymb}

\newcommand{\de}{\partial}

\newcommand{\eq}[2]{\begin{equation} \label{#1} #2 \end{equation}}

\newcommand{\diag}{\textrm{diag}}
\newcommand{\etal}{{\em et al.}}

\begin{document}

\title{Optical analogue of relativistic Dirac solitons in binary waveguide arrays}
\author{Truong X. Tran$^{1,2}$, Stefano Longhi$^{3}$ and Fabio Biancalana$^{2,4}$}
\affiliation{$^{1}$Dept. of Physics, Le Quy Don University, 236 Hoang Quoc Viet Street, 10000 Hanoi, Vietnam\\
$^{2}$Max Planck Institute for the Science of Light, G\"{u}nther-Scharowsky str. 1, 91058 Erlangen, Germany\\
$^{3}$Dept. of Physics, Politecnico di Milano and Istituto di Fotonica e Nanotecnologie del Consiglio Nazionale delle Ricerche, Piazza L. da Vinci 32, I-20133 Milano, Italy\\
$^{4}$School of Engineering and Physical Sciences, Heriot-Watt University, EH14 4AS Edinburgh, UK}
\date{\today}

\begin{abstract}
We study analytically and numerically an optical analogue of Dirac solitons in binary waveguide arrays in presence of Kerr nonlinearity. Pseudo-relativistic soliton solutions of the coupled-mode equations describing dynamics in the array are analytically derived. We demonstrate that with the found soliton solutions, the coupled mode equations can be converted into the nonlinear relativistic 1D Dirac equation. This paves the way for using binary waveguide arrays as a classical simulator of quantum nonlinear effects arising from the Dirac equation, something that is thought to be impossible to achieve in conventional (i.e. linear) quantum field theory.

\end{abstract}
\maketitle

\paragraph{Introduction ---} Waveguide arrays have been used intensively to simulate the evolution of a nonrelativistic quantum mechanical particle in a periodic potential. Many fundamental phenomena in nonrelativistic classical and quantum mechanics such as Bloch oscillations \cite{pertsch,morandotti2}, Zener tunneling \cite{ghulinyan,trompeter}, optical dynamical localization \cite{longhi0}, and Anderson localization in disordered lattices \cite{lahini} have been simulated both theoretically and experimentally with waveguide arrays. In a recent study it was shown that, rather surprisingly, most of nonlinear fiber optics features (such as resonant radiation and soliton self-wavenumber shift) can also take place in specially excited arrays \cite{tranresonant}. Recently, binary waveguide arrays (BWAs)  have also been used to mimic relativistic phenomena typical of quantum field theory, such as Klein tunneling \cite{longhi1,dreisow2}, the {\em  Zitterbewegung} (trembling motion of a free Dirac electron) \cite{longhi2,dreisow}, and fermion pair production \cite{longhi3}, which are all based on the properties of the Dirac equation \cite{zeuner}. Although there is as yet no evidence for fundamental quantum nonlinearities, nonlinear versions of the Dirac equation have been studied since a long time. One of the earlier extensions was provided by Heisenberg \cite{Heisenberg57} in the context of field theory and was motivated by the question of mass. In the quantum mechanical context, nonlinear Dirac equations have been used as effective
theories in atomic, nuclear and gravitational physics \cite{NLD1,NLD2,NLD3,NLD4} and, more recently, in the study of ultracold atoms \cite{haddad,haddad11}. To this regard, BWAs  can offer a rather unique model system to simulate nonlinear extensions of the Dirac equation when probed at high light intensities.  The discrete gap solitons in BWAs in the {\em classical} context have been investigated both numerically \cite{sukhorukov1,sukhorukov2,conforti11} and experimentally \cite{morandotti}. In particular, in Ref. \cite{sukhorukov2} soliton profiles with even and odd symmetry were numerically calculated and a scheme with two Gaussian beams, which are tuned to the Bragg angle with opposite inclinations, was proposed to efficiently generate gap solitons. In Ref. \cite{morandotti} solitons were experimentally observed when the inclination angle of an input beam is slightly above the Bragg angle.

Inspired by the importance of BWAs as a classical simulator for relativistic quantum phenomena, and also by past achievements in the  investigation of discrete gap solitons in BWAs, in this Letter we present analytical soliton solutions of the discrete coupled-mode equations (CMEs) for a BWA and construct Dirac solitons of a nonlinear relativistic 1D Dirac equation  in the quasicontinuous limit. This paves the way for using BWAs to simulate nonlinear extensions  of the Dirac equation that violate Lorentz invariance \cite{note}, as well as other solitonic and nonsolitonic effects of nonlinear Dirac equations.

\paragraph{Analytical soliton solutions ---}
Light propagation in a discrete, periodic binary array of Kerr nonlinear waveguides can be described, in the continuous-wave regime, by the following dimensionless CMEs \cite{sukhorukov1,longhi1}:
\eq{CWCM}{i\frac{da_{n}(z)}{dz} = -\kappa[a_{n+1}(z)+ a_{n-1}(z)] + (-1)^{n} \sigma a_{n} -  \gamma |a_{n}(z)|^{2}a_{n}(z),}
where $a_{n}$ is the electric field amplitude in the $n$th waveguide, $z$ is the longitudinal spatial coordinate, $2\sigma$ and $\kappa$ are the propagation mismatch and the coupling coefficient between two adjacent waveguides of the array, respectively, and $\gamma$ is the nonlinear coefficient of waveguides which is positive for self-focusing, but negative for self-defocusing media. For simplicity, here we suppose all waveguides have the same nonlinear coefficient, but even if these nonlinear coefficients are different (provided they are comparable), then analytical soliton solutions shown later will not be changed, because as explained later, one component of solitons is much weaker than both unity and other component, and thus one can eliminate the nonlinear term associated with this weak soliton component. In the dimensionless form, in general, one can normalize variables in the above equation such that $\gamma$ and $\kappa$ are equal to unity. However, throughout this Letter we will keep these parameters explicitly in Eqs. (\ref{CWCM}). Before proceeding further, it is helpful to analyze the general properties of the general solutions of Eqs. (\ref{CWCM}). First of all, let us assume that $(a_{2n},a_{2n-1})^{T} = i^{2n}(\varphi_{2n},\varphi_{2n-1})^{T}$ is one solution of Eqs. (\ref{CWCM}) with $\varphi_{2n}$ and $\varphi_{2n-1}$ being appropriate functions. In this case, if we change the sign of $\gamma$ while keeping the other two parameters constant, one can easily show that a new solution of Eqs. (\ref{CWCM}) will be $(a_{2n},a_{2n-1})^{T} = i^{2n}(\varphi^{*}_{2n-1},\varphi^{*}_{2n})^{T}$, where $*$ denotes the complex conjugation. Secondly, if the sign of $\sigma$ is changed while other parameters kept constant, then a new solution of Eqs. (\ref{CWCM}) will be $(a_{2n},a_{2n-1})^{T} = i^{2n}(\varphi_{2n-1},\varphi_{2n})^{T}$. Of course, when $\sigma$ changes sign, we still have the same physical system, but with a shift of the wavenumber position $n$ in Eqs. (\ref{CWCM}) by one. The above simple rules allow us to quickly find other solutions and their symmetries if one particular solution is known, as will be shown later.

In the specific case when all three parameters $\gamma, \kappa$, and $\sigma$ are all kept positive, we look for analytical solutions of motionless solitons of Eqs. (\ref{CWCM}) in the following form:
\eq{solutionform}{\small \left[\begin{array}{cc} a_{2n}(z)
\\ a_{2n-1}(z) \end{array}\right] = \left[\begin{array}{cc} i^{2n} d \frac{2}{n_{0}} \mathrm{sech}(\frac{2n}{n_{0}}) e^{ifz}
\\ -i^{2n-1}b \mathrm{sech}(\frac{2n-1}{n_{0}}) \mathrm{tanh}(\frac{2n-1}{n_{0}}) e^{ifz}
\end{array}\right],} where $n_{0}\in\mathbb{R}$ characterizes the beam width (i.e. the average number of waveguides on which the beam extends), and coefficients $b, d$ and $f$ are still unknown. In the system without any loss or gain of energy (i.e., when $\kappa, \sigma$ and $\gamma$ are all real), the coefficient $f$ must also be real, but $b$ and $d$ can be complex in general. Inserting the ansatz (\ref{solutionform}) into Eqs. (\ref{CWCM}), assuming {\em a priori}  that the component $a_{2n-1}$ is much weaker than both unity and the other component $a_{2n}$, such that one can eliminate the nonlinear term for $a_{2n-1}$, and also assuming that the quasicontinuous limit is valid (i.e. $n_{0} $ is large enough), after some lengthy algebra one gets:
\begin{eqnarray}
fd &=& \kappa bi - \sigma d, \label{algebra1} \\
i\kappa b &=& 2\gamma|d|^{2}d/n^2_{0}, \label{algebra2} \\
fb &=& \sigma b + 4d\kappa i/n^2_{0}. \label{algebra3}
\end{eqnarray}
Extracting $f$ and $b$ from Eq. (\ref{algebra1}) and Eq. (\ref{algebra2}), respectively, then inserting them into Eq. (\ref{algebra3}) we will get one quadratic equation for $d^{2}$, and thus can find the values for $b, d$ and $f$. Note that one needs to keep only solutions which satisfy the above assumption that $|a_{2n-1}| \ll |a_{2n}|$. The final solution in the case when $\gamma, \kappa, \sigma>0$ is:
\eq{solitonsolution}{\small \left[\begin{array}{cc} a_{2n}(z)
\\ a_{2n-1}(z) \end{array}\right] = \left[\begin{array}{cc} i^{2n} \frac{2\kappa}{n_{0}\sqrt{\sigma\gamma}} \mathrm{sech}(\frac{2n}{n_{0}}) e^{iz(\frac{2\kappa^{2}}{n^{2}_{0}\sigma}-\sigma)}
\\ i^{2n} \frac{2\kappa^{2}}{n^{2}_{0}\sigma\sqrt{\sigma\gamma}} \mathrm{sech}(\frac{2n-1}{n_{0}}) \mathrm{tanh}(\frac{2n-1}{n_{0}}) e^{iz(\frac{2\kappa^{2}}{n^{2}_{0}\sigma}-\sigma)}
\end{array}\right].}

It is worth mentioning that the analytical soliton solution in the form of Eqs. (\ref{solitonsolution}) is derived under two conditions: (i) the beam must be large enough such that one can operate in the quasicontinuous limit instead of the discrete one; and (ii) $n_{0}|\sigma| \gg 2\kappa$. The latter condition is easily satisfied if (i) is held true and if $\sigma$ is comparable to $\kappa$ \cite{dreisow}. If condition (ii) is not valid, one can still easily get the analytical solution for $b, d$ and $f$ from Eqs. (\ref{algebra1}) - (\ref{algebra3}), but they are a bit cumbersome and for brevity we do not show it here. The solution in form of Eqs. (\ref{solitonsolution}) represents a one-parameter family of discrete solitons in BWAs where the beam width parameter $n_{0}$  can be arbitrary, provided that $n_{0} \gtrsim4$, a surprisingly small number for the quasi continuous approximation to be valid.
\begin{figure}[htb]
  \centering \includegraphics[width=0.45\textwidth]{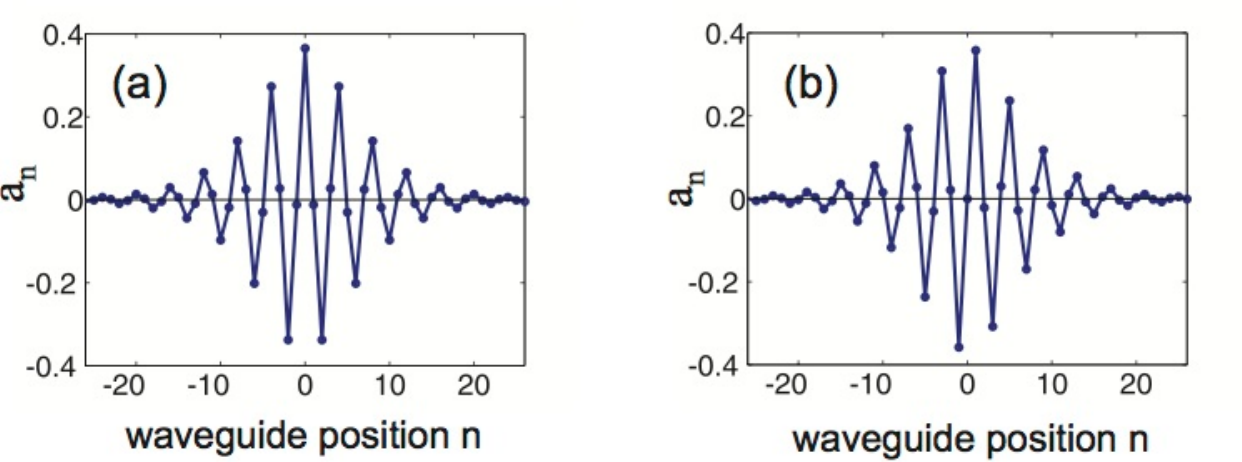}
\caption{\small{(Color online) Discrete soliton profiles (a,b) for even and odd symmetry, respectively. Full circles mark the field amplitudes across the BWA. Parameters in (a): $\kappa$ =1; $\gamma$ =1; $\sigma$ = 1.2; and $n_{0} = 5$. After getting the even symmetry profile in (a), we construct the odd profile in (b) by switching the sign of $\sigma$ and following the symmetry transformations explained in the text.}}
  \label{fig1}
\end{figure}

\begin{figure}[htb]
  \centering \includegraphics[width=0.45\textwidth]{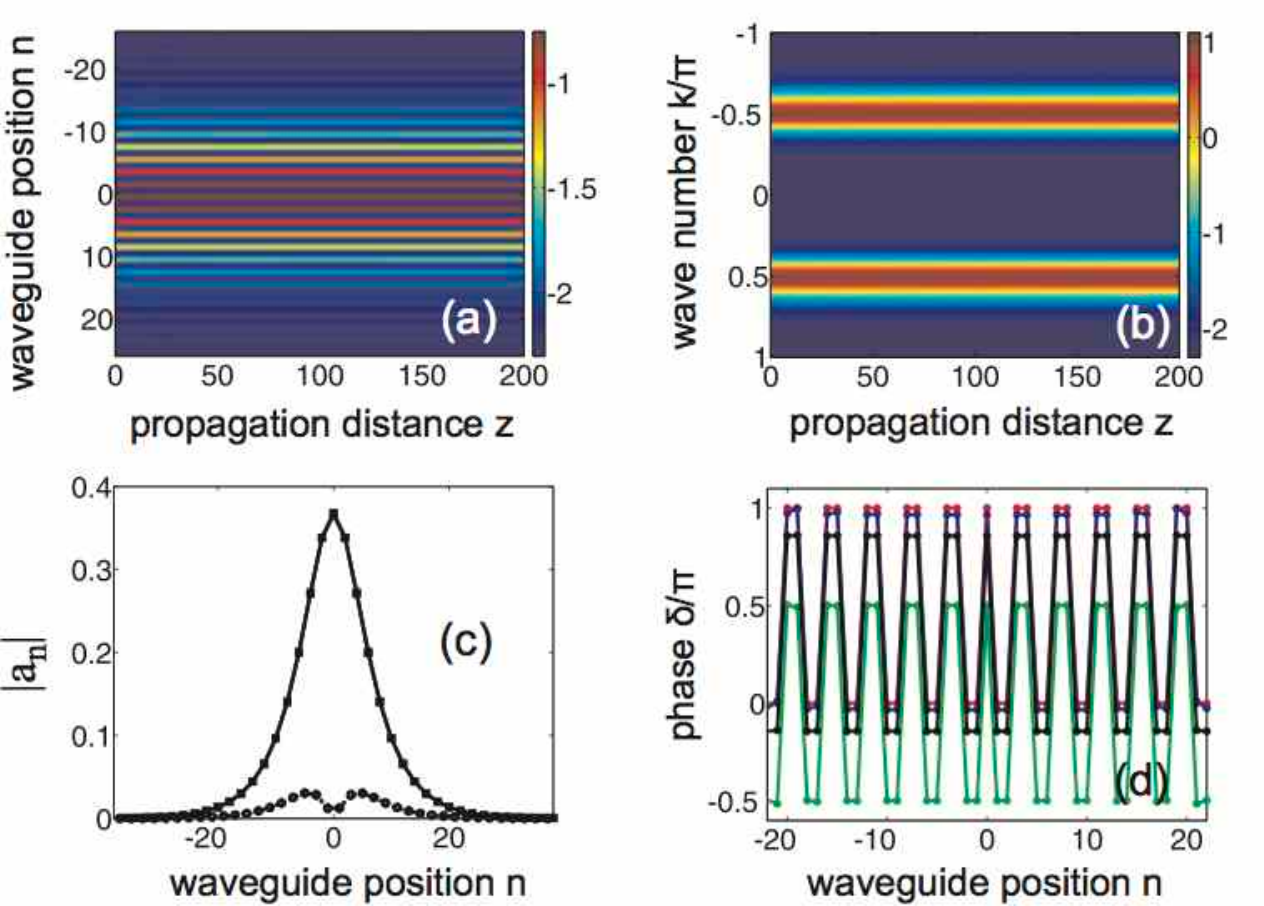}
\caption{\small{(Color online) (a,b) Soliton propagation in the $(n,z)$-plane (a) and its Fourier transform in the $(k,z)$-plane (b) with even symmetry profile at the input. (c) Absolute values of the field amplitudes for intense ($|a_{2n}|$ with solid line and square markers) and weak ($|a_{2n-1}|$ with dashed-dotted curves and round markers) components of soliton at four different values of $z=0$ (red curves); $50$ (blue curves); $140$ (green curves); and $200$ (black curves). Soliton profile is so well preserved that all these curves just stay on top of each other and one can see only the output black curves. (d) Phase pattern $\delta/\pi$ of soliton profiles at four above values of $z$. Colors of curves in (d) have the same meaning as in (c). Parameters: $\kappa$ =1; $\gamma$ =1; $\sigma$ = 1.2; and $n_{0} = 5$. All contour plots are shown in logarithmic scale.}}
  \label{fig2}
\end{figure}

In Fig. \ref{fig1}(a) we plot the soliton profile with even symmetry calculated by using Eqs. (\ref{solitonsolution}) at $z$ = 0 with full circles marking the field amplitudes across BWAs, for the parameters given in the caption. Note that soliton profile in Fig. \ref{fig1}(a) consists of two components: one strong component $a_{2n}$ and another much weaker component $a_{2n-1}$ [see also Fig. \ref{fig2}(c)]. Once we get the soliton solution in Fig. \ref{fig1}(a), we can construct another soliton solution of the same physical system by changing the sign of $\sigma$ and following the rules explained in the previous section. In that way we obtain the odd symmetry soliton profile depicted in Fig. \ref{fig1}(b). It is important to mention that in the case of self-focusing media ($\gamma >$ 0), for both even and odd symmetries the strong component is always located at waveguides with larger propagation constant [channels with $+|\sigma|$ in Eqs. (\ref{CWCM})], whereas the weak component is located at waveguides with smaller propagation constant [channels with $-|\sigma|$ in Eqs. (\ref{CWCM})]. We are also able to construct the soliton solutions for the self-defocusing media which also possess soliton solutions with even and odd symmetries. The only difference with the self-focusing media is that now the strong (weak) component is located at waveguides with smaller (larger) propagation constant.

\paragraph{Soliton propagation and generation --}
Equation (\ref{solitonsolution}) and the associated solutions obtained by  the above symmetry transformations provide the analytical forms of the two discrete gap soliton branches numerically found in \cite{sukhorukov2}. We note that the propagation constant $f$ of the two solitons, given by $f=-\sigma+2 \kappa^2 / (n_0^2 \sigma)$,  falls in the minigap of the superlattice, near the edge of the lower miniband (because $2 \kappa^2 / (n_0^2 \sigma) \ll \sigma$), and thus they are expected to be stable \cite{sukhorukov2}.    As an example, in Fig. \ref{fig2}(a) we show the soliton propagation along $z$ as obtained by numerically solving Eqs. (\ref{CWCM}) with an input soliton taken from Eqs. (\ref{solitonsolution}) at $z=0$, demonstrating that the soliton profile is well preserved during propagation. Parameters used for Fig. \ref{fig2} are the same as in Fig. \ref{fig1}(a). The evolution of the Fourier transform of the field $a_n$ in Fig. \ref{fig2}(a) along z is shown in Fig. \ref{fig2}(b) where the wavenumber $k$ represents the phase difference between adjacent waveguides. Due to the periodic nature of BWAs, within the coupled mode approximation, it suffices to investigate $k$ in the first Brillouin zone  $-\pi\leq k \leq \pi$ \cite{lederer}. One very important feature of the wavenumber evolution in Fig. \ref{fig2}(b) is the fact that there are two components of wavenumber centered at $k = \pm \pi/2$ which correspond to two Bragg angles \cite{dreisow} with opposite inclinations. These two wavenumber components are generated at the input and preserve their shapes during propagation along $z$. This feature of $k$ indicates that the soliton operates in the region where CMEs could potentially be converted into the relativistic Dirac equations describing the evolution of a freely moving relativistic particle \cite{longhi2,dreisow}. We will come back to this important point again later. Figure \ref{fig2}(c) shows the two components of the soliton profile at odd and even waveguide position $n$. The strong component with solid curves and square markers represents the field profile $|a_{2n}|$ at even waveguide positions, whereas the weak component with dashed-dotted curves and round markers represents the field profile $|a_{2n-1}|$ at odd waveguide positions. Field profiles in Fig. \ref{fig2}(c) are taken at four values of propagation distance $z=0$ (red curves); $50$ (blue curves); $140$ (green curves); and $200$ (black curves) -- only the black curves are actually visible since the the profile is perfectly preserved during propagation with a very high precision. The soliton profile also perfectly preserves its phase pattern across the array [Fig. \ref{fig2}(d)]. From Eqs. (\ref{solitonsolution}), one can easily see that as the waveguide position variable $n$ runs, the phase pattern of the soliton must be periodic as follows: $\delta_{n} = ... (\rho, \rho), (\rho + \pi, \rho + \pi), (\rho, \rho)...$ where $\rho$ also changes with $z$. This pattern is only broken at the soliton center point where the function $tanh$ in Eqs. (\ref{solitonsolution}) changes its sign. This phase pattern is shown in Fig. \ref{fig2}(d) where different colors with meanings as in Fig. \ref{fig2}(c) depict pattern at different values of $z$. The sequence in the phase is important because it allows us to convert Eqs. (\ref{CWCM}) into the nonlinear Dirac equation as we shall show shortly. Note that the soliton whose propagation is shown in Fig. \ref{fig2} is the one with even symmetry in  Fig. \ref{fig1} (a). Our simulations similarly show that the profile of soliton with odd symmetry in Fig. \ref{fig1} (b) is also well preserved during propagation, and we have checked that this is true even in presence of quite a strong numerical noise, demonstrating the robustness and the stability of our solutions.

\begin{figure}[htb]
  \centering \includegraphics[width=0.45\textwidth]{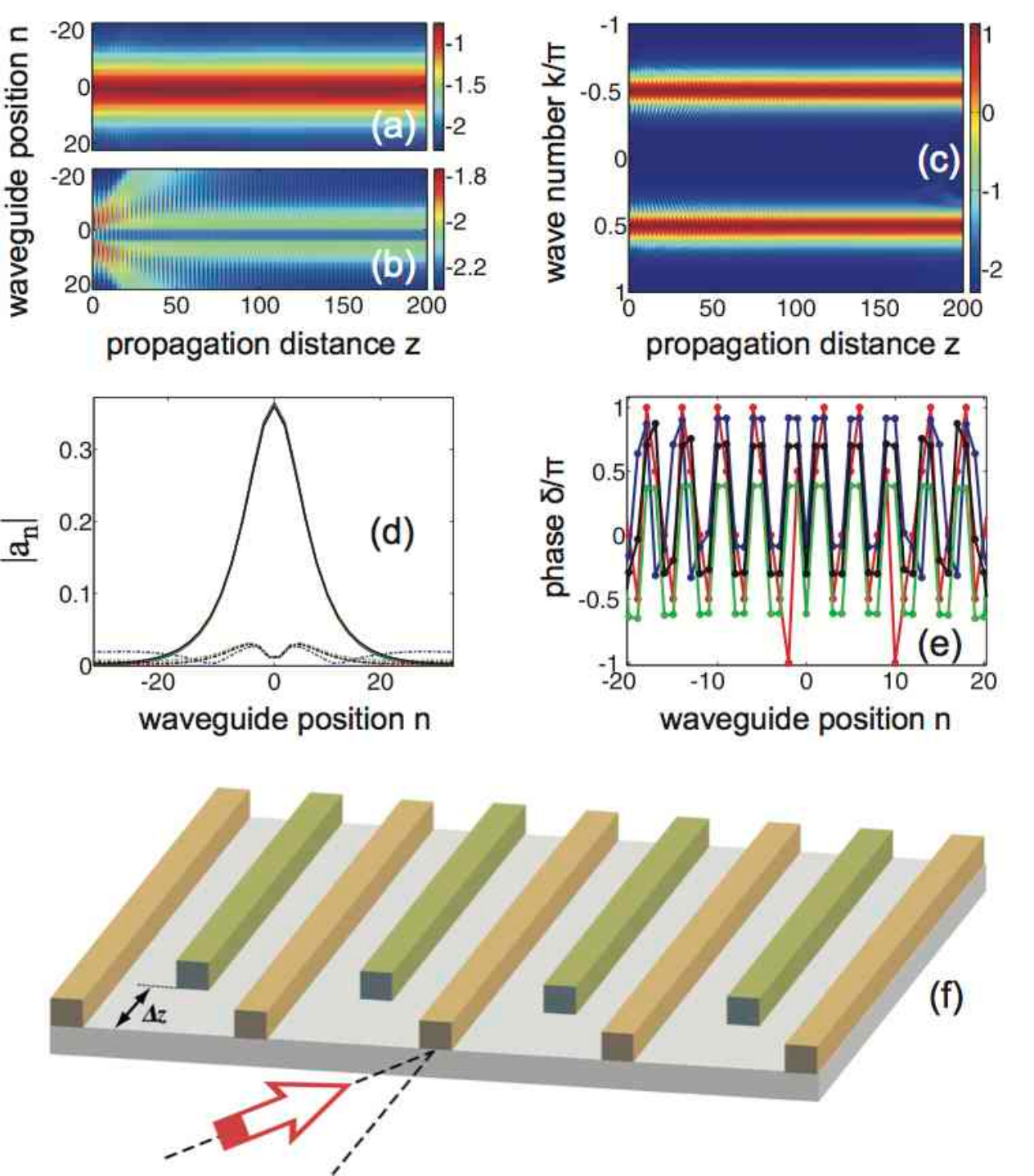}
\caption{\small{(Color online) (a,b) Propagation in the $(n,z)$-plane of the even and odd components of the beam with initial phase difference equal to $\pi/2$ between adjacent waveguides. (c) Fourier transform of field amplitudes in the $(k,z)$-plane. (d) Absolute values of the field amplitudes for intense component $|a_{2n}|$ with solid curves and weak component $|a_{2n-1}|$ with dashed-dotted curves at four different values of $z$ = 0 (red curves); 50 (blue curves); 140 (green curves); and 200 (black curves). (e) Phase pattern $\delta/\pi$ of field amplitudes for the same values of $z$ as in (d). Colors of curves in (e) have the same meaning as in (d). (f) Scheme of the BWA structure for generating discrete solitons. Parameters: $\kappa$ =1; $\gamma$ =1; $\sigma$ = 1.2; and $n_{0} = 5$.}}
  \label{fig3}
\end{figure}

Although the soliton solutions given by Eqs. (\ref{solitonsolution}) are exact, it is important to consider the possibility to generate the new gap solitons by an input beam with a simpler (and more experimentally accessible) profile. Due to the wavenumber structure shown in Fig. \ref{fig2}(b), one can interpret the soliton as a combination of two beams launched under two Bragg angles with opposite tilts $k = \pm \pi/2$, similarly to what was suggested in Ref. \cite{sukhorukov2}. Here we propose to generate the soliton by an input with a simple phase pattern where the phase difference between adjacent waveguides is equal to $\pi/2$ across the array. The input condition is taken to be $A_n = a_{n}\mathrm{exp}(in\pi/2)$ where $a_{n}$ is given by Eqs. (\ref{solitonsolution}) at $z$ = 0, but without the term $i^{2n}$. Note that, since $|a_{2n-1}| \ll |a_{2n}|$, this input condition can be approximately achieved by exciting the BWA with a broad beam tilted at the Bragg angle, with the odd waveguides in the structure being realized at some spatial delay $\Delta z$ inside the sample (so as they are not excited at the input plane); see the scheme shown in Fig. \ref{fig3}(f). In the linear regime, the beam broadens and undergoes {\it Zitterbewegung} \cite{longhi2,dreisow}, whereas in the nonlinear regime soliton formation is expected to take place with suppression of both beam broadening and {\it Zitterbewegung}. This is clearly shown in Fig. \ref{fig3}, which indicates the formation of the soliton during propagation with parameters as in Fig. \ref{fig2}. The evolution of field profiles $|a_{2n}|$ and $|a_{2n-1}|$ at even and odd waveguide positions is depicted in Fig. \ref{fig3}(a) and \ref{fig3}(b), respectively. The evolution of the Fourier transform of the field $a_n$ of Fig. \ref{fig3}(a,b) along $z$ is shown in Fig. \ref{fig3}(c). One can see that the strong component $a_{2n}$ in Fig. \ref{fig3}(a) does not change much during propagation, whereas the weak component $a_{2n-1}$ in Fig. \ref{fig3}(b) is dramatically altered during propagation. As seen from Fig. \ref{fig3}(b), at the beginning of the propagation the beam undergoes the {\em  Zitterbewegung}. After reaching $z\simeq$ 70, the profile $|a_{2n-1}|$ becomes stable.  Figure \ref{fig3}(d) shows the strong component $|a_{2n}|$ of the soliton profile with solid curves and the weak component $|a_{2n-1}|$ with dashed-dotted curves. As in Fig. \ref{fig2}(c,d) field profiles are taken at four values of propagation distance $z$ = 0 (red curves); 50 (blue curves); 140 (green curves); and 200 (black curves). One can also see that the strong component $|a_{2n}|$ is stable, whereas the weak component first gets distorted (see blue and green curves), but eventually the output curve (black color) relaxes to the input curve (red color). Figure \ref{fig3}(e) depicts the phase pattern of the field amplitudes across the array calculated at different $z$ with corresponding colors as in Fig.
 \ref{fig3}(d). At the input (red curve) we have the phase difference equal to $\pi/2$ between adjacent waveguides, but this phase pattern quickly transforms into the phase pattern of the soliton solution given by Eqs. (\ref{solitonsolution}), i.e., $\delta_{n} = ... (\rho, \rho), (\rho + \pi, \rho + \pi), (\rho, \rho)...$ [see blue, green and black curves in Fig. \ref{fig3}(e)]. Therefore, here one can make a local conclusion: a beam with the intensity profiles of soliton solution given by Eqs. (\ref{solitonsolution}), but with phase difference equal to $\pi/2$ between adjacent waveguides will first undergo {\em Zitterbewegung}, but eventually its intensity profile and phase pattern will relax to the ones of the soliton solution given by Eqs. (\ref{solitonsolution}).

\paragraph{Dirac solitons ---}
As mentioned in the introduction, BWAs have been used to mimic phenomena in both nonrelativistic and relativistic quantum mechanics. To the best of our knowledge so far all these phenomena which have been simulated by BWAs are linear. In this section we will report on the simulation of nonlinear relativistic Dirac solitons in BWAs. As shown in \cite{longhi2,dreisow} linear CMEs [Eqs. (\ref{CWCM})] for a beam with phase difference equal to $\pi/2$ can be converted into the linear one-dimensional relativistic Dirac equation (DE). Note that Eqs. (\ref{CWCM}) can be converted into the DE only for beams with special phase patterns; for instance, at normal beam incidence Eqs. (\ref{CWCM}) can not be converted into the DE. It turns out that with the soliton solution given by Eqs. (\ref{solitonsolution}), one can also successfully convert Eqs. (\ref{CWCM}) into the nonlinear relativistic Dirac equation (NDE). Thus, one can use BWAs to mimic the relativistic Dirac solitons, and soliton solutions in BWAs given by Eqs. (\ref{solitonsolution}) can be used to construct directly the Dirac soliton. Although the solution of Eqs. (\ref{solitonsolution}) does not possess the phase difference equal to $\pi/2$ between adjacent waveguides [see Fig. \ref{fig2}(d)], the fact that it exhibits two wavenumbers $k=\pm \pi/2$ [see Fig. \ref{fig2}(b)] gives us some hope that the NDE can also be obtained in this case. Indeed, this is the case as shown below. In general, suppose that $[a_{2n}(z), a_{2n-1}(z)]^{T}$ = $i^{2n}[g(2n,z),q(2n-1,z)]^{T}$, where the two functions $g$ and $q$ are smooth and their derivatives $\de_{n}g$ and $\de_{n}q$ exist in the quasicontinuous limit [Eqs. (\ref{solitonsolution}) satisfy these requirements]. After setting $\Psi_{1}(n) = (-1)^{n}a_{2n}$ and $\Psi_{2}(n) = i(-1)^{n}a_{2n-1}$, and following the standard approach developed in \cite{longhi2,dreisow} we can introduce the continuous transverse coordinate $\xi \leftrightarrow n$ and the  two-component spinor $\Psi(\xi,z)$ = $(\Psi_{1},\Psi_{2})^{T}$ which satisfies the 1D NDE:
\eq{diracequation}{i\de_{z}\Psi = -i\kappa\alpha\de_{\xi}\Psi + \sigma\beta\Psi - \gamma G,} where the nonlinear terms $G \equiv (|\Psi_{1}|^{2}\Psi_{1},|\Psi_{2}|^{2}\Psi_{2})^{T}$; $\beta=\diag(1,-1)$ is the Pauli matrix $\sigma_{z}$; and $\alpha$ is the Pauli matrix $\sigma_{x}$ with diagonal elements equal to zero, but off-diagonal elements equal to unity. Note that Eq. (\ref{diracequation}) is identical to the DE obtained in \cite{longhi2,dreisow} with the only difference that now we have the nonlinear term $G$ in Eq. (\ref{diracequation}). Similar soliton solutions have been found for the NDE in Ref. \cite{nogami}, but with different and more complicated kind of nonlinearity, in the context of quantum field theory. Note that the nonlinearity that we have in Eq. (\ref{diracequation}) violates  Lorentz invariance \cite{note}, and is similar to that of the Dirac equations in Bose-Einstein condensates \cite{haddad}. Using the soliton solution given by Eqs. (\ref{solitonsolution}) and the above relation between $a_{n}$ and $\Psi$ one can easily obtain the Dirac soliton solution of Eq. (\ref{diracequation}) as follows:
\eq{Diracsoliton}{\small \left[\begin{array}{cc} \Psi_{1}(\xi,z)
\\ \Psi_{2}(\xi,z) \end{array}\right] = \left[\begin{array}{cc} \frac{2\kappa}{n_{0}\sqrt{\sigma\gamma}} \mathrm{sech}(\frac{2\xi}{n_{0}}) e^{iz(\frac{2\kappa^{2}}{n^{2}_{0}\sigma}-\sigma)}
\\ i \frac{2\kappa^{2}}{n^{2}_{0}\sigma\sqrt{\sigma\gamma}} \mathrm{sech}(\frac{2\xi-1}{n_{0}}) \mathrm{tanh}(\frac{2\xi-1}{n_{0}}) e^{iz(\frac{2\kappa^{2}}{n^{2}_{0}\sigma}-\sigma)}
\end{array}\right].}
The above solution is obtained for $\sigma >0$ and $\gamma >0$. One can use the symmetry properties of Eqs. (\ref{CWCM}) to construct other Dirac soliton solutions of  Eq. (\ref{diracequation}), with different sign combinations between $\sigma$ and $\gamma$.
The expressions given by Eq. (\ref{Diracsoliton}) give the main result of this Letter, and the only physically realizable way that we are aware of to produce and observe Dirac solitons with a table-top experiment. In future investigations we are planning to carefully study the dynamics and the stability of Dirac solitons in BWAs, on which we will report in a separate publication.

\paragraph{Conclusions ---} In this Letter we have provided analytical expressions for the non-moving  gap solitons in BWAs and shown their connection to Dirac solitons in a nonlinear extension of the relativistic 1D Dirac equation describing the dynamics of a freely moving relativistic particle. Our results suggest that BWAs can be used as a classical simulator to investigate relativistic Dirac solitons, enabling to realize an experimentally accessible model system of quantum nonlinearities that have been so far a subject of speculation in the foundation of quantum field theories. The analysis of analogue of quantum field theory effects as those ones described in this Letter is applicable to virtually any nonlinear discrete periodic system supporting solitons, either classical or quantum, therefore making our results very general and of relevance to different systems beyond optics, such as ultracold atoms in optical lattices and trapped ions where analogs of linear relativistic effects, such as {\em Zitterbewegung}, have been studied and observed \cite{uff1,uff2,uff3}

This work is supported by the German Max Planck Society for the Advancement of Science (MPG).

\end{document}